\documentclass[12pt]{article}
\usepackage{amsmath, amsthm, amssymb, amsfonts, enumerate,bm,url}
\usepackage[figuresright]{rotating}
\usepackage{lscape}
\usepackage{multirow}
\usepackage{natbib}

\usepackage{algorithm,algpseudocode} 
\usepackage{booktabs}

\begin{document}
\newcommand{\abs}[1]{\left\vert#1\right\vert}
\newcommand{\set}[1]{\left\{#1\right\}}
\newcommand{\eps}{\varepsilon}
\newcommand{\To}{\rightarrow}
\newcommand{\inv}{^{-1}}
\newcommand{\ihat}{\hat{\imath}}
\newcommand{\var}{\mbox{Var}}
\newcommand{\sd}{\mbox{SD}}
\newcommand{\cov}{\mbox{Cov}}
\newcommand{\f}{\frac}
\newcommand{\fI}[1]{\frac{1}{#1}}
\newcommand{\what}[1]{\widehat{#1}}
\newcommand{\hhat}[1]{\what{\what{#1}}}
\newcommand{\wtilde}[1]{\widetilde{#1}}
\newcommand{\bdot}{\bm{\cdot}}
\newcommand{\Th}{\theta}
\newcommand{\qmq}[1]{\quad\mbox{#1}\quad}
\newcommand{\qm}[1]{\quad\mbox{#1}}
\newcommand{\mq}[1]{\mbox{#1}\quad}
\newcommand{\tr}{\mbox{tr}}
\newcommand{\logit}{\mbox{logit}}
\newcommand{\noi}{\noindent}
\newcommand{\bni}{\bigskip\noindent}
\newcommand{\bul}{$\bullet$ }
\newcommand{\bias}{\mbox{bias}}
\newcommand{\conv}{\mbox{conv}}
\newcommand{\spn}{\mbox{span}}
\newcommand{\colspace}{\mbox{colspace}}
\newcommand{\mF}{\mathcal{F}}
\newcommand{\mI}{\mathcal{I}}
\newcommand{\mJ}{\mathcal{J}}
\newcommand{\mL}{\mathcal{L}}
\newcommand{\mP}{\mathcal{P}}
\newcommand{\mS}{\mathcal{S}}
\newcommand{\mT}{\mathcal{T}}
\newcommand{\mX}{\mathcal{X}}
\newcommand{\bbR}{\mathbb{R}}
\newcommand{\fwer}{\mbox{FWE}}
\newcommand{\fweI}{FWE$_I$}
\newcommand{\fweII}{FWE$_{II}$}
\newcommand{\fdr}{\mbox{FDR}}
\newcommand{\fnr}{\mbox{FNR}}
\newcommand{\pfdr}{\mbox{pFDR}}
\newcommand{\pfnr}{\mbox{pFNR}}
\newcommand{\mte}{\mbox{MTE}}
\newcommand{\vphi}{\varphi}
\newcommand{\Bern}{\mbox{Bern}}
\newcommand{\oN}{\overline{N}}

%  Mike's commands
\newcommand{\onem}{\left\{ 1,\ldots,m\right\} }
\newcommand{\st}{\text{ s.t. }}
\newcommand{\gaussian}{\mbox{Gaussian}}
\newcommand{\fdp}{\mbox{FDP}}
\newcommand{\fnp}{\mbox{FNP}}
\newcommand{\poisson}{\mbox{Poisson}}
\newcommand{\binomial}{\mbox{Binomial}}

\newtheorem{theorem}{Theorem}[section]
\newtheorem{corollary}{Corollary}[section]
\newtheorem{conjecture}{Conjecture}[section]
\newtheorem{proposition}{Proposition}[section]
\newtheorem{lemma}{Lemma}[section]
\newtheorem{definition}{Definition}[section]
\newtheorem{example}{Example}[section]
\newtheorem{remark}{Remark}[section]

\title{{\bf\Large Sequential FDR and  pFDR control under arbitrary dependence, with application to pharmacovigilance database monitoring}}

\author{\textsc{Michael Hankin}\footnote{Mothball Labs}\; and \textsc{Jay Bartroff}\footnote{Corresponding author. University of Texas at Austin. Email: \texttt{bartroff@austin.utexas.edu}}\\
%\small{Department of Mathematics, University of Southern California, Los Angeles, California, USA}\\ 
%\small{\textsf{\{bartroff, jinlinso\}@usc.edu}}
}  

\date{}
\maketitle

\abstract{We propose sequential multiple testing procedures which control the false discovery rate~(FDR) or the positive false discovery rate~(pFDR) under arbitrary dependence between the data streams. This is accomplished by ``optimizing'' an upper bound on these error metrics for a class of step down sequential testing procedures.  Both open-ended and truncated versions of these sequential procedures are given, both being able to control both the type~I multiple testing metric (FDR or pFDR) at specified levels, and the former being able to control both the type~I and type~II (e.g., FDR \emph{and} the false nondiscovery rate, FNR). In simulation studies, these procedures provide 45-65\% savings in average sample size over their fixed-sample competitors.  We illustrate our procedures on drug data from the United Kingdom's Yellow Card Pharmacovigilance Database.} 

\section{Introduction}

The majority of the procedures proposed in the statistics literature for multiple testing for fixed sample size or sequential data can be bifurcated into either step up or step down procedures.  Step up procedures decide whether to accept or reject null hypotheses in order of increasing significance, whereas step down procedures operate in the reverse order. For testing $J\ge 2$ null hypotheses, the Benjamini-Hochberg \citeyearpar[][hereafter BH]{Benjamini95} procedure is step up, and is known to control the false discovery rate (FDR) at the nominal level~$\alpha$ under independence of $p$-values, and at no more than the inflated level~$C\alpha $, where $C=C(J):=\sum_{j=1}^J 1/j$,  under arbitrary dependence among $p$-values, with $C\approx \log(J)$ for large $J$. Although \citet{Benjamini01} showed that FDR can be controlled at the uninflated level~$\alpha$ under certain types of positive dependence, \citet{Guo08} showed that the FDR bound $C\alpha$ is sharp for the step up BH procedure in the sense that there exists a joint distribution of $p$-values for which the FDR equals this. Analogous results were established by \citet{Bartroff18} and \citet{Bartroff20} about step up procedures for sequential data.

In this sense, the ``worst case'' performance of step up procedures is essentially completely understood, at least without imposing additional restrictions or assumptions on the joint distribution of $p$-values.  So in this paper we turn out attention to studying and controlling the worst case performance of step down procedures, in the setting of sequential data, i.e., $J$ data streams or arbitrary dependence. More specifically, we give  a step down procedure for sequential data and the smallest possible constant~$D<C$ whose FDR is bounded above by $D\alpha$ where, again, $\alpha$ is the nominal FDR level.
 
 \citet{Hart97} used linear programming to find a sharp upper bound on the probability of rejection for a class of multiple testing procedures in terms of the true, marginal c.d.f.s of the $p$-values. Calculating their bound  therefore requires knowledge of at least the marginals of the true joint distribution.  \citet{Guo08} also used optimization but to find an upper bound on the false discovery rate~(FDR) in terms of a step down procedure's step values, that holds under arbitrary joint distribution of the $p$-values. By choosing the step values to control the upper bound, they produce a step down procedure whose FDR control holds under arbitrary joint distributions.  The current paper extends this line of research in the following ways. First, by extending to the sequential setting an FDR controlling procedure without requiring or imposing assumptions on the joint distribution of the $p$-values; the open-ended versions of these sequential procedures allow the possibility of simultaneously controlling both type~I and type~II  error metrics (e.g., FDR \emph{and} the false nondiscovery rate, FNR). Secondly, by removing an implicit assumption on the $p$-values in Guo and Rao's proof that is stronger than validity of $p$-values and applies to both the sequential and fixed-sample settings; see Appendix~\ref{sec:inf.pf} for details. Finally, by extending this methodology to positive false discovery rate (pFDR) and its type~II analog.
 
 The rest of this paper is organized as follows. After giving the notational set up in Section~\ref{sec:setup}, Section~\ref{sec:untrunc} covers untruncated (i.e., open-ended) sequential procedures and records the FDR/FNR- and pFDR/pFNR-controlling procedures in Theorems~\ref{thm:inf.fdr} and \ref{thm:inf.pfdr}, respectively, as special cases of a generic untruncated procedure given in Section~\ref{sec:gen.D}. Truncated sequential procedures are handled analogously in Section~\ref{sec:trunc} with the corresponding FDR- and pFDR-controlling procedures given in Theorems~\ref{thm:fin.fdr} and \ref{thm:fin.pfdr}, respectively.  To aide the reader, Table~\ref{tab:thms} is a ``directory'' of these theorems and cases. In Section~\ref{sec:sims} we present simulation studies on Binomial and Poisson data of the proposed sequential procedures and compare with their fixed sample counterparts, and in Section~\ref{sec:Drug-Data} we apply our proposed procedures to data from the UK's Yellow Card Pharmacovigilance Database.  We end with conclusions and discussion in Section~\ref{sec:conc}.  Proofs are given in the Appendix.
 
 \begin{table}[H]
 \centering
 \caption{Directory of this paper's error control theorems}
 \label{tab:thms}
 \bigskip
\begin{tabular}{c|c|c}
 Error Metrics & Untruncated, Type I and II Controlling& Truncated, Rejective \\
\hline 
$\fdr/\fnr$ & Theorem~\ref{thm:inf.fdr}& Theorem~\ref{thm:fin.fdr}\\
\hline 
$\pfdr/\pfnr$ & Theorem~\ref{thm:inf.pfdr}& Theorem~\ref{thm:fin.pfdr} \\
\end{tabular}
\end{table}

\subsection*{Dedication}
This paper is dedicated to the memory of Prof.~Tze Leung Lai, who influenced and inspired the authors directly and indirectly. While the current paper is not a direct offshoot of one of Prof.~Lai's works, it overlaps with some of Lai's most active research areas including sequential hypothesis testing \citep{Lai81,Lai88,Lai01}, adaptive designs \citep{Lai79,Lai87}, multiple comparisons \citep{Lai00,Bartroff10e}, optimization \citep{Han06}, quality control \citep{Lai95}, longitudinal data \citep{Lai06b}, and biomedical applications \citep{Bartroff13}, to name a few references.

\section{Set up}\label{sec:setup}

\subsection{Data Streams, Hypotheses, and Error Metrics}
Our general form of data streams and hypotheses is as follows.  Assume that there are $J\ge 2$ data streams
\begin{align}
\mbox{Data stream $1$:}\quad&X_1^{(1)},X_2^{(1)},\ldots\nonumber\\
\mbox{Data stream $2$:}\quad&X_1^{(2)},X_2^{(2)},\ldots\label{streams}\\
\vdots&\nonumber\\
\mbox{Data stream $J$:}\quad&X_1^{(J)},X_2^{(J)},\ldots.\nonumber
\end{align}
In order to implement our proposed procedures, the marginal type~I and II error probabilities of component test statistics on each stream will need to be controlled at certain levels; see \eqref{typeI}-\eqref{typeII} below.  Beyond that, we make no assumptions about the dimension  of the sequentially-observed data $X_n^{(j)}$, which may themselves be vectors of varying size, nor about the dependence structure of within-stream data $X_n^{(j)}, X_m^{(j)}$ or between-stream data $X_n^{(j)}, X_m^{(j')}$ ($j\ne j'$). In particular there can be arbitrary ``overlap'' between data streams, an extreme case being that all the data streams are the same, which is equivalent to testing multiple hypotheses about a single data source. For any positive integer $j$ let $[j]=\{1,\ldots,j\}$. For each data stream $j\in[J]$, assume that there is a parameter vector $\theta^{(j)}\in\Theta^{(j)}$ determining that distribution of the stream $X_1^{(j)}, X_2^{(j)},\ldots$, and it is desired to test a null hypothesis~$H^{(j)}$ versus an alternative hypothesis~$G^{(j)}$, where $H^{(j)}$ and $G^{(j)}$ are disjoint subsets of the parameter space~$\Theta^{(j)}$ containing $\theta^{(j)}$. It is \emph{not} required that $H^{(j)}\cup G^{(j)}=\Theta^{(j)}$, e.g., one-sided alternatives or separated hypotheses are possible. The null hypothesis~$H^{(j)}$ is considered \textit{true} if $\theta^{(j)}\in H^{(j)}$, and \textit{false} if $\theta^{(j)}\in G^{(j)}$. The global parameter $\theta=(\theta^{(1)},\ldots,\theta^{(J)})$ is the concatenation of the individual parameters and is contained in the global parameter space $\Theta=\Theta^{(1)}\times\cdots\times \Theta^{(J)}$.

The general notation~\eqref{streams} includes fully-sequential sampling where the streamwise sample sizes may take any value $1,2,\ldots$ \textit{ad infinitum}, but other sampling set ups fit this as well including group sequential, truncated, and even fixed sample size testing. For example,  the $n$th ``observation''~$X_n^{(j)}$ in the $j$th stream may actually be the $n$th group $X_n^{(j)}=(X_{n,1}^{(j)},\ldots,X_{n,\ell}^{(j)})$ of size $\ell$. Moreover, the group size~$\ell$ may vary with $n$ and may even be data-dependent, e.g., determined by some type of adaptive sampling. Similarly, truncated sequential (or group sequential) sampling can be implemented for the $j$th stream by defining $X_n^{(j)}=\emptyset$ for all $n>\overline{N}^{(j)}$ for some stream-specific truncation point $\overline{N}^{(j)}$, or globally for all streams by replacing statements like ``for some $n$'' in what follows with ``for some $n\le \overline{N}$,'' for some global truncation point $\overline{N}$. One may represent a fixed sample size in the $j$th stream in this way by taking $\overline{N}^{(j)}=1$, or in all streams with $\overline{N}=1$.

For any multiple testing procedure under consideration, let $V$ denote the number of true null hypotheses it rejects (i.e., the number of false positives), $W$ the number of false null hypotheses it accepts (i.e., the number of false negatives), and $R$ the number of null hypotheses it rejects. The number of null hypotheses accepted is therefore $J-R$. Under the true value of the parameter~$\theta$, the false discovery and non-discovery rates \citep[][FDR, FNR]{Benjamini95} are
\begin{equation*}
\fdr=\fdr(\theta)=E_\theta\left(\frac{V}{R\vee 1}\right),\quad \fnr=\fnr(\theta)=E_\theta\left(\frac{W}{(J-R)\vee 1}\right),
\end{equation*}
where $x\vee y=\max\{x,y\}$. Similarly, the positive false discovery rate~pFDR \citep{Storey02} and its type~II analog, the positive false nondiscovery rate~pFNR, are defined as
\begin{equation}\label{pFDR.defs}
\pfdr=\pfdr(\theta)=E_\theta\left(\left.\frac{V}{R}\right|R\ge 1\right)\qmq{and} \pfnr=\pfnr(\theta)=E_\theta\left(\left.\frac{W}{J-R}\right|J-R\ge 1\right).
\end{equation} 

\subsection{Test Statistics and Critical Values}\label{sec:stats}
The building blocks of the sequential procedures defined below are $J$ individual sequential test statistics $\{\Lambda^{(j)}(n)\}_{j\in[J],\; n\ge 1}$, where $\Lambda^{(j)}(n)$ is the statistic for testing $H^{(j)}$ vs.\ $G^{(j)}$ based on the data $X_1^{(j)},X_2^{(j)},\ldots,X_n^{(j)}$ available  from the $j$th stream at time $n$.  For example, in parametric settings $\Lambda^{(j)}(n)$ may be a sequential  log (generalized) likelihood ratio statistic for testing $H^{(j)}$ vs.\ $G^{(j)}$. Like the fixed sample size procedures mentioned above, our sequential procedures will utilize \textit{step values} (or \textit{step value vectors}) which are $J$-long vectors of non-decreasing values in $[0,1]$, such as
\begin{equation}\label{UD.values}
\bm{\alpha}=(\alpha_1,\ldots, \alpha_J)\qmq{with} 0<\alpha_1\le\alpha_2\le \ldots\le\alpha_J\le 1.
\end{equation} 
Given two step value vectors~$\bm{\alpha}$ and $\bm{\beta}$, we say that a sequential test statistic~$\{\Lambda^{(j)}(n)\}_{n\ge 1}$ is \textit{implemented with step values $\bm{\alpha}, \bm{\beta}$} if there are critical values~$\{A_k^{(j)}, B_k^{(j)}\}_{k\in[J]}$ such that
\begin{align}
P_{\theta^{(j)}}(\Lambda^{(j)}(n)\ge B_k^{(j)}\;\mbox{some $n$,}\; \Lambda^{(j)}(n')>A_1^{(j)}\;\mbox{all $n'<n$})&\le \alpha_k\qmq{for all}\theta^{(j)}\in H^{(j)}\label{typeI}\\
 P_{\theta^{(j)}}(\Lambda^{(j)}(n)\le A_k^{(j)}\;\mbox{some $n$,}\; \Lambda^{(j)}(n')<B_1^{(j)}\;\mbox{all $n'<n$})&\le \beta_k\qmq{for all}\theta^{(j)}\in G^{(j)}\label{typeII}
\end{align} 
for all $k\in[J]$. The critical values $A_1^{(j)}, B_1^{(j)}$ are simply the critical values for the sequential test that samples until $\Lambda^{(j)}(n)\not\in (A_1^{(j)}, B_1^{(j)})$, and this test has type~I and II error probabilities bounded above by $\alpha_1$ and $\beta_1$, respectively.  The values $B_k^{(j)}$, $k\in[J]$, are then such that the similar sequential test with critical values $A_1^{(j)}$ and $B_k^{(j)}$ has type~I error probability $\alpha_k$, which is just a restatement of \eqref{typeI}, with an analogous statement holding for critical values $A_k^{(j)}$ and $B_1^{(j)}$, type~II error probability $\beta_k$, and \eqref{typeII}.  We say that the test statistics~$\{\Lambda^{(j)}(n)\}_{j\in[J],\; n\ge 1}$ for all the streams are \textit{implemented with step values $\bm{\alpha}, \bm{\beta}$} if they are for each stream. In all commonly-encountered testing situations there are standard sequential statistics which can be implemented with given step values $\bm{\alpha}, \bm{\beta}$. \citet{Bartroff14b,Bartroff20} and \citet{Bartroff18}  give examples.

Without loss of generality we assume that, for each $j\in[J]$,  
\begin{gather}
A_1^{(j)}\le A_2^{(j)}\le\ldots\le A_J^{(j)}\le B_J^{(j)}\le B_{J-1}^{(j)}\le\ldots\le B_1^{(j)}, \label{AB.mono}\\
\mbox{$A_k^{(j)}=A_{k+1}^{(j)}$ if and only if  $\beta_k=\beta_{k+1}$}\label{A.mono}\\
\mbox{$B_k^{(j)}=B_{k+1}^{(j)}$ if and only if  $\alpha_k=\alpha_{k+1}$.}\label{B.mono}
\end{gather}
The sequential multiple testing procedures proposed below will involve ranking the test statistics associated  with different data streams, which may be on completely different scales in general, so for each stream $j$ we introduce a \textit{standardizing function} $\vphi^{(j)}(\cdot)$ which will be applied to the statistic $\Lambda^{(j)}(n)$ before ranking. The standardizing functions  $\vphi^{(j)}$ can be any increasing functions such that $\vphi^{(j)}(A_k^{(j)})$ and $\vphi^{(j)}(B_k^{(j)})$ do not depend on $j$, and we let
\begin{equation}\label{ab.def}
a_k=\vphi^{(j)}(A_k^{(j)})\qmq{and}b_k=\vphi^{(j)}(B_k^{(j)}),\quad j,k\in[J],
\end{equation} denote these common values.  Given critical values $\{A_k^{(j)},B_k^{(j)}\}_{j,k\in[J]}$ satisfying \eqref{typeI}-\eqref{typeII}, one may choose arbitrary values $\{a_k,b_k\}_{k\in[J]}$ satisfying the same monotonicity conditions as the $\{A_k^{(j)},B_k^{(j)}\}$ according to \eqref{A.mono}-\eqref{B.mono} and then define the standardizing functions $\vphi^{(j)}(\cdot)$ to be increasing, piecewise linear functions satisfying \eqref{ab.def}. For example, if all the $\alpha_k$ are distinct and the $\beta_k$ are distinct then a simple choice for the $\{a_j,b_j\}$ are the integers
$$a_1=-J,\quad a_2=-J+1,\quad \ldots,\quad a_J=-1, \quad b_J=1, \quad b_{J-1}=2,\quad \ldots,\quad b_1=J.$$
In any case, the assumptions on the critical values and standardizing functions imply that the $a_k$ must be nondecreasing and the $b_k$ nonincreasing.
Finally, we denote $\wtilde{\Lambda}^{(j)}(n)= \varphi^{(j)}(\Lambda^{(j)}(n))$ and then \eqref{typeI}-\eqref{typeII} can be written
\begin{align}
P_{\theta^{(j)}}(\wtilde{\Lambda}^{(j)}(n)\ge b_k\;\mbox{some $n$,}\; \wtilde{\Lambda}^{(j)}(n') > a_1\;\mbox{all $n'<n$})&\le \alpha_k\qmq{for all}\theta^{(j)}\in H^{(j)}\label{typeI.stand}\\
 P_{\theta^{(j)}}(\wtilde{\Lambda}^{(j)}(n)\le a_k\;\mbox{some $n$,}\; \wtilde{\Lambda}^{(j)}(n')< b_1\;\mbox{all $n'<n$})&\le \beta_k \qmq{for all}\theta^{(j)}\in G^{(j)},\label{typeII.stand}
\end{align} for all $j,k\in[J]$.

The theorems in this paper will give upper bounds on the FDR and pFDR, and their type~II versions, for sequential procedures implemented with arbitrary step values~$\bm{\alpha}, \bm{\beta}$. In practice we recommend using some of the commonly used values such as those recommended by \citet{Benjamini95} or \citet{Benjamini99}.
For desired value~$q_1\in(0,1)$ of the type~I error metric (FDR or pFDR), the former are
\begin{equation}\label{BH.step}
\bm{\alpha}_j=\bm{\alpha}(q_1)_j :=q_1 j/J\qmq{for}j\in[J],
\end{equation} and the latter are
\begin{equation}\label{BL.step}
\bm{\alpha}_j=\bm{\alpha}(q_1)_j=1-\left(1-\left(1\wedge \frac{q_1J}{J-j+1}\right)\right)^{1/(J-j+1)}\qmq{for}j\in[J].
\end{equation} 
Here $x\wedge y=\min\{x,y\}$. We recommend using the same expressions for $\bm{\beta}=\bm{\beta}(q_2)$ for desired value~$q_2\in(0,1)$ of the type~II error metric (FNR or pFNR). In our examples below we will refer to \eqref{BH.step} and \eqref{BL.step} as the BH and BL step values, respectively.

\subsection{Simple hypotheses and Wald approximations}
In practice, many testing situations can be reduced to, or approximated by, testing simple vs.\ simple hypotheses.  \citet[][Section~1]{Muller07} point out that testing a battery of simple-vs.-simple hypothesis tests has been the standard set up for use of FDR in the literature. In this section we show how to construct the test statistics~$\Lambda^{(j)}(n)$ and critical values $A_s^{(j)}, B_s^{(j)}$ satisfying \eqref{typeI}-\eqref{typeII} for any data stream $j$ such that $H^{(j)}$ and $G^{(j)}$ are both simple hypotheses.  In this case the test statistic~$\Lambda^{(j)}(n)$ can  be taken to be log-likelihood ratio (density under $G^{(j)}$ divided by the density under $H^{(j)}$) yielding the sequential probability ratio test \citep[SPRT; see][]{Chernoff72}. See \citet[][Section~5.1]{Bartroff20} for a more formal description of hypotheses and parameter space in this set up. In the single null hypothesis testing set up, the SPRT samples until $\Lambda^{(j)}(n)\not\in(A, B)$ where the critical values $A,B$ are chosen so that
\begin{align}
P_{H^{(j)}}(\Lambda^{(j)}(n)\ge B\;\mbox{some $n$,}\; \Lambda^{(j)}(n')>A\;\mbox{all $n'<n$})&\le \alpha\label{SPRT-typeI}\\
P_{G^{(j)}}(\Lambda^{(j)}(n)\le A\;\mbox{some $n$,}\; \Lambda^{(j)}(n')<B\;\mbox{all $n'<n$})&\le\beta\label{SPRT-typeII}
\end{align}
 for desired type I and II error probabilities $\alpha$ and $\beta$. The most common way of choosing the critical values~$A, B$ is to use the closed-form \emph{Wald-approximations} 
 \begin{equation}\label{myAB}
A=A_W(\alpha,\beta) :=\log\left(\frac{\beta}{1-\alpha}\right)+\rho,\quad B=B_W(\alpha,\beta):=\log\left(\frac{1-\beta}{\alpha}\right)-\rho
\end{equation} 
for which it is assumed that $\alpha+\beta\le 1$ and $\rho\ge 0$ is a fixed adjustment term to account for the test statistic's excess over the boundary upon stopping. See \citet[][Section~3.3.1]{Hoel71} for a derivation of Wald's \citeyearpar{Wald47} original $\rho=0$ case and, based on Brownian motion approximations, \citet[][p.~50 and Chapter~X]{Siegmund85} derives the value $\rho=.583$ which has been used to improve the approximation for continuous random variables. With our multiple testing examples below we recommend using Siegmund's $\rho=.583$.

In order to apply the above to the $j$th  stream in the multiple testing set up for given step value vectors~$\bm{\alpha}$ and $\bm{\beta}$, a slight extension of \citet[][Theorem~5.1]{Bartroff20} shows that choosing
\begin{equation}\label{W4SvS}
A_k^{(j)}=A_W(\wtilde{\alpha}_k,\beta_k),\quad B_k^{(j)}=B_W(\alpha_k,\wtilde{\beta}_k)\qmq{for} k\in[J],
\end{equation} where
\begin{equation}\label{surr.AB4W}
\wtilde{\alpha}_k:=\frac{\alpha_1(1-\beta_k)}{1-\beta_1}\qmq{and}\wtilde{\beta}_k := \frac{\beta_1(1-\alpha_k)}{1-\alpha_1} \qmq{for} k\in[J],
\end{equation}
satisfies \eqref{typeI}-\eqref{typeII} up to Wald's approximations, and that $\wtilde{\alpha}_k+\beta_k\le 1$ and $\alpha_k+\wtilde{\beta}_k\le 1$ for all $k\in[J]$. Although, strictly speaking, the inequalities \eqref{typeI}-\eqref{typeII} will only be guaranteed to hold approximately using these approximations,  \citet{Hoel71} show that the actual type I and II error probabilities can only exceed the nominal values by a negligibly small amount in the worst case, and the difference approaches $0$ for small nominal values, which is relevant in the present multiple testing situation where we will utilize fractions of $\alpha$ and $\beta$. Alternatives to the Wald approximations in the simple vs.\ simple set up are Monte Carlo or to replace the terms in \eqref{myAB} by the values $\log \beta$ and $\log(1/\alpha)$, respectively, for which \eqref{SPRT-typeI}-\eqref{SPRT-typeII} hold conservatively \citep[see][]{Hoel71} and proceed similarly. 

\section{Procedures controlling type~I and II error metrics}\label{sec:untrunc}

\subsection{The Generic Sequential Step Down Procedure}\label{sec:gen.D}

Next we review the generic sequential step down procedure, defined in \citet{Bartroff18}. Versions of this procedure, implemented with certain step values~$\bm{\alpha}, \bm{\beta}$, will produce our proposed type~I and II FDR and pFDR controlling sequential procedures.  We assume that test statistics are implemented with $\bm{\alpha}, \bm{\beta}$, and the critical values referred to are those satisfying \eqref{typeI}-\eqref{typeII}.

We describe the procedure in terms of stages of sampling, between which reject/accept decisions are made. Let $\mJ_i\subseteq[J]$ ($i=1,2,\ldots$) denote the index set of the \emph{active} data streams, those whose corresponding null hypothesis $H^{(j)}$ has been neither accepted nor rejected yet at the beginning of the $i$th stage of sampling, and $n_i$ denote the cumulative sample size of any active test statistic up to and including the $i$th stage. The total number of null hypotheses that have been rejected (resp.\ accepted) at the beginning of the $i$th stage is denoted by $r_i$ (resp.\ $c_i$). Accordingly, set $\mJ_1=[J]$, $n_0=0$, $r_1=c_1 =0$. Let $|\cdot|$ denote set cardinality. Then the $i$th stage of sampling ($i=1,2,\ldots$) of the \textit{Generic Sequential Step Down Procedure} implemented with step values $\bm{\alpha}, \bm{\beta}$ proceeds as follows.

\begin{enumerate}
\item\label{sample-step} Sample the active streams $\{X_n^{(j)}\}_{j\in\mJ_i, \; n>n_{i-1}}$ until $n$ equals
\begin{equation}\label{cont-samp}n_i=\inf\left\{n>n_{i-1}: \wtilde{\Lambda}^{(j)}(n)\not\in (a_{c_i+1}, b_{r_i+1}) \qmq{for some} j\in\mJ_i \right\}.\end{equation}
\item\label{step:ord} Order the active test statistics
\begin{equation*}
\wtilde{\Lambda}^{(j(n_i,1))}(n_i)\le \wtilde{\Lambda}^{(j(n_i,2))}(n_i)\le \ldots\le \wtilde{\Lambda}^{(j(n_i,|\mJ_i|))}(n_i),\end{equation*} where $j(n_i,\ell)$ denotes the index of the $\ell$th ordered active statistic at the end of stage~$i$.
\item  
\begin{enumerate}
\item\label{rej-step} If the upper boundary in \eqref{cont-samp} has been crossed, $\wtilde{\Lambda}^{(j)}(n_i)\ge b_{r_i+1}$ for some $j\in\mJ_i$, then reject the $t_i\ge 1$ null hypotheses 
\begin{equation*}
H^{(j(n_i,|\mJ_i|))}, H^{(j(n_i,|\mJ_i|-1))}, \ldots, H^{(j(n_i,|\mJ_i|-t_i+1))}, 
\end{equation*}
where
\begin{equation*}
t_i=\max\left\{t\in [|\mJ_i|]: \wtilde{\Lambda}^{(j(n_i,\ell))}(n_i)\ge b_{r_i+|\mJ_i|-\ell+1}\qmq{for all}\ell=|\mJ_i|-t+1,\ldots,|\mJ_i| \right\},
\end{equation*} 
and set $r_{i+1}=r_i+t_i$. Otherwise set $r_{i+1}=r_i$.

\item\label{acc-step} If the lower boundary in \eqref{cont-samp} was crossed, $\wtilde{\Lambda}^{(j)}(n_i)\le a_{c_i+1}$ for some $j\in\mJ_i$, then accept the $t_i'\ge 1$ null hypotheses $$H^{(j(n_i,1))}, H^{(j(n_i,2))}, \ldots, H^{(j(n_i,t_i'))},$$ where 
\begin{equation*}
t_i'=\max\left\{t\in[|\mJ_i|] : \wtilde{\Lambda}^{(j(n_i,\ell))}(n_i)\le a_{c_i+\ell} \qmq{for all}\ell=1,\ldots,t\right\},
\end{equation*} and set $c_{i+1}=c_i+t_i'$. Otherwise set $c_{i+1}=c_i$.
\end{enumerate}

\item\label{stop-step} Stop if there are no remaining active hypotheses, $r_{i+1}+c_{i+1}=J$.  Otherwise, let $\mJ_{i+1}$ be the indices of the remaining active hypotheses and continue on to stage~$i+1$.
\end{enumerate} 

Thus the procedure samples all active data streams until at least one of the active null hypotheses can be accepted or rejected, indicated by the stopping rule~\eqref{cont-samp}. At that point, step down rejection/acceptance rules are used in steps \ref{rej-step}/\ref{acc-step} to reject/accept some active null hypotheses. After updating the list of active hypotheses, the process is repeated until no active hypotheses remain.

\subsection{Untruncated sequential procedures controlling FDR/FNR or pFDR/pFNR}

For $m\in\{0,1,\ldots,J\}$  and  step values~$\bm{\alpha}$, define
\begin{align} 
D(\bm{\alpha},m)&=m\left(\sum_{j=1}^{J-m+1}\frac{\alpha_{j}-\alpha_{j-1}}{j}+(J-m)\sum_{j=J-m+2}^J \frac{\alpha_{j}-\alpha_{j-1}}{j(j-1)}\right),\label{Dm}\\
D(\bm{\alpha})&=\max_{0\le m\le J} D(\bm{\alpha}, m).\label{D}
\end{align}
Our main results in this section, Theorems~\ref{thm:inf.fdr} and \ref{thm:inf.pfdr}, show that the generic sequential step down procedure applied with step values $\bm{\alpha}$ have error metrics (e.g., FDR) bounded above by $D(\bm{\alpha},m_0)$ where $m_0$ is the number of true null hypotheses, which is in turn bounded above by $D(\bm{\alpha})$. Since $m_0$ is usually unknown in practice, this latter bound is typically more useful in practice, and the former is primarily of interest for theoretical understanding of the procedure, although there are situations wherein the number of true nulls is assumed to be known \citep[e.g.,][]{Song17,He21}. It follows from $D(\bm{\alpha},m)$ being linear in $\bm{\alpha}$ that 
\begin{equation}\label{D.lin}
D(c\bm{\alpha}) = cD(\bm{\alpha})\qm{for any $c>0$}.
\end{equation}
 Hence an error metric of a procedure utilizing step values $\bm{\alpha}$ being bounded above by $D(\bm{\alpha})$ is equivalent to saying that, for a desired value~$q\in(0,1)$ of the error metric, the procedure utilizing step values $\wtilde{\bm{\alpha}}=q\bm{\alpha}/D(\bm{\alpha})$ has the error metric bounded above by $q$.  

The proofs of Theorems~\ref{thm:inf.fdr} and \ref{thm:inf.pfdr} are delayed until the Appendix.

\begin{theorem} \label{thm:inf.fdr} The generic sequential step down procedure implemented with step values $\bm{\alpha}, \bm{\beta}$ satisfies
\begin{equation}\label{inf.fdr.m.bd}
\fdr(\theta)\le D(\bm{\alpha},m_0)\le D(\bm{\alpha})\qmq{and} \fnr(\theta)\le D(\bm{\beta},m_1)\le D(\bm{\beta})\qmq{for all}\theta\in\Theta 
\end{equation}
regardless of the dependence between data streams, where $D$ is as defined in \eqref{Dm}-\eqref{D} and $m_0$ and $m_1$ are the numbers of true and false null hypotheses, respectively. In particular, given $q_1, q_2\in (0,1)$, the generic sequential step down procedure implemented with step values $\wtilde{\bm{\alpha}}=q_1\bm{\alpha}/D(\bm{\alpha})$, $\wtilde{\bm{\beta}}=q_2\bm{\beta}/D(\bm{\beta})$ satisfies
\begin{equation}\label{inf.fdr.alpha.bd}
\fdr(\theta)\le q_1\qmq{and} \fnr(\theta)\le q_2\qmq{for all}\theta\in\Theta 
\end{equation}
regardless of the dependence between data streams.
\end{theorem}

For the next theorem, recall that $R$ denotes the number of null hypotheses rejected by the procedure in question.

\begin{theorem} \label{thm:inf.pfdr} Let $D$ be as defined in \eqref{Dm}-\eqref{D} and $m_0$ and $m_1$ denote the numbers of true and false null hypotheses, respectively. Fix arbitrary $\theta\in\Theta$.
\begin{enumerate}
\item If $\gamma_1,\gamma_2>0$ are values such that the generic sequential step down procedure implemented with step values $\bm{\alpha}, \bm{\beta}$ satisfies
\begin{equation}\label{gamma.lower}
P_\theta(R>0)\ge \gamma_1\qmq{and} P_\theta(R<J)\ge \gamma_2,
\end{equation}
then this procedure satisfies
\begin{equation}\label{inf.pfdr.m.bd}
\pfdr(\theta) \le \frac{D(\bm{\alpha},m_0)}{P_\theta(R>0)}\le  \frac{D(\bm{\alpha})}{\gamma_1}\qmq{and} \pfnr(\theta)\le \frac{D(\bm{\beta},m_1)}{P_\theta(R<J)}\le  \frac{D(\bm{\beta})}{\gamma_2}
\end{equation}
regardless of the dependence between data streams. 

\item\label{part:inf.pfdr.q} In particular, given $q_1, q_2\in (0,1)$,  the generic sequential step down procedure implemented with step values $\wtilde{\bm{\alpha}}=q_1\gamma_1\bm{\alpha}/D(\bm{\alpha})$, $\wtilde{\bm{\beta}}=q_2\gamma_2\bm{\beta}/D(\bm{\beta})$ satisfies
\begin{equation}\label{inf.pfdr.alpha.bd}
\pfdr(\theta)\le q_1\qmq{and} \pfnr(\theta)\le q_2
\end{equation}
regardless of the dependence between data streams, if $\gamma_1,\gamma_2>0$ satisfy \eqref{gamma.lower} for this procedure.
 
\item\label{part:inf.pfdr.gij} For $j\in[J]$ let
\begin{align}
\gamma_{1j}&= P_{\theta^{(j)}}(\wtilde{\Lambda}^{(j)}(n)\ge b_1\;\mbox{some $n$,}\; \wtilde{\Lambda}^{(j)}(n') > a_J\;\mbox{all $n'<n$}),\label{g1j.def}\\
\gamma_{2j}&= P_{\theta^{(j)}}(\wtilde{\Lambda}^{(j)}(n)\le a_1\;\mbox{some $n$,}\; \wtilde{\Lambda}^{(j)}(n')< b_J\;\mbox{all $n'<n$}).\nonumber
\end{align}
If $\max_{j\in[J]}\gamma_{1j}>0$ and $\max_{j\in[J]}\gamma_{2j}>0$, then \eqref{gamma.lower}-\eqref{inf.pfdr.m.bd}  hold with $\gamma_i=\max_{j\in[J]}\gamma_{ij}$, $i=1,2$.
\end{enumerate}

\end{theorem}

\section{Truncated, rejective procedures}\label{sec:trunc}
In this section we describe versions of our procedures which only stop early to reject (rather than accept) null hypotheses  and thus only explicitly control the corresponding type~I multiple testing error rate (FDR or pFDR), recorded in Theorems~\ref{thm:fin.fdr}  and \ref{thm:fin.pfdr}.  This setting naturally corresponds with having a streamwise maximum sample size (or ``truncation point'')~$\overline{N}$ which we assume throughout this section. For this reason we refer to them as ``truncated, rejective'' versions of the procedures.  These procedures may be preferable in certain situations such as when (a) a null hypothesis being true represents the system being ``in control'' and therefore continued sampling (rather than stopping) is desirable, (b) there is a maximum sample size imposed on the data streams possible preventing simultaneous achievement of the nominal error bounds \eqref{typeI}-\eqref{typeII}, or (c) the type~II multiple testing error rate (e.g, FNR)~$q_2$ is not well-motivated.  In any of theses cases, one may prefer to drop the requirement that the type~II multiple testing error rate be strictly controlled at an arbitrary level~$q_2$ and use one of the rejective procedures which, roughly speaking, are similar but ignore the lower stopping boundaries $A_k^{(j)}$. On the other hand, if $q_2$ is not well motivated but the statistician prefers early stopping under the null hypotheses, we encourage the use of one of our procedures in Section~\ref{sec:untrunc} with both early stopping to reject and accept null hypotheses, and treat~$q_2$ as  a parameter to be chosen to give a procedure with other desirable operating characteristics, such as expected total or streamwise maximum sample size.

\subsection{Set up and critical values}\label{sec:trunc.setup}

The set up for rejective procedures requires a few modifications. Let the data streams $X_n^{(j)}$, test statistics $\Lambda^{(j)}(n)$, and parameters $\theta^{(j)}$ and $\theta$ be as in Section~\ref{sec:setup}. Since only the type~I multiple testing error rate, FDR or pFDR, will be explicitly controlled we only require specification of null hypotheses $H^{(j)}\subseteq\Theta^{(j)}$ and not of alternative hypotheses $G^{(j)}$. We assume a streamwise maximum sample size~$\overline{N}$ for each stream, but with only notational changes, what follows could be formulated by stream-specific truncation points $\{\overline{N}^{(j)}\}_{j\in [J]}$ or with sample sizes other than  $1,\ldots,\overline{N}$. 

Given step values $\bm{\alpha}$, we adapt our definition from Section~\ref{sec:stats} of the test statistics~$\{\Lambda^{(j)}(n)\}_{j, n}$  being \textit{implemented with step values $\bm{\alpha}$} to this truncated, rejective setting if, for all $j,k\in[J]$, the critical values $B_1^{(j)},\ldots,B_J^{(j)}$ satisfy
\begin{equation*}
P_{\theta^{(j)}}\left( \Lambda^{(j)}(n)\ge B_k^{(j)}\;\mbox{for some $n\le\overline{N}$}\right)\le \alpha_k\qmq{for all} \theta^{(j)} \in H^{(j)},
\end{equation*} as well as \eqref{AB.mono} and \eqref{B.mono} without loss of generality. We let the standardizing functions~$\varphi^{(j)}$ be any increasing functions such that $b_k=\varphi^{(j)}(B_k^{(j)})$ does not depend on $j$, and $\wtilde{\Lambda}^{(j)}(n)=\varphi^{(j)}(\Lambda^{(j)}(n))$ denote the standardized statistics.

In the next section we give the truncated, rejective version of the generic step down procedure, and then in Theorems~\ref{thm:fin.fdr}  and \ref{thm:fin.pfdr} state their FDR and pFDR controlling properties. The proofs are similar to the proofs of Theorems~\ref{thm:inf.fdr} and \ref{thm:inf.pfdr} and are sketched in the Appendix.
 
\subsection{The generic rejective sequential  step down procedure}\label{sec:RD.gen}
With the notation of Section~\ref{sec:gen.D}, the $i$th stage ($i=1,2,\ldots$) of the \textit{Generic Rejective Sequential Step Down Procedure} with step values $\bm{\alpha}$ proceeds as follows.

\begin{enumerate}
\item Sample the active streams $\{X_n^{(j)}\}_{j\in\mJ_i, \; n>n_{i-1}}$ until $n$ equals
\begin{equation}\label{cont-samp.rej}n_i=\oN\wedge \inf\left\{n>n_{i-1}: \wtilde{\Lambda}^{(j)}(n) \ge b_{r_i+1}\qmq{for some} j\in\mJ_i \right\}.\end{equation}

\item If $n_i=\oN$ and no test statistic has crossed the critical value in \eqref{cont-samp.rej}, accept all active null hypotheses and terminate the procedure.  Otherwise, proceed to Step~\ref{step:ord.rej}.

\item\label{step:ord.rej} Order the active test statistics
\begin{equation*}
\wtilde{\Lambda}^{(j(n_i,1))}(n_i)\le \wtilde{\Lambda}^{(j(n_i,2))}(n_i)\le \ldots\le \wtilde{\Lambda}^{(j(n_i,|\mJ_i|))}(n_i)
\end{equation*} 
and reject the $t_i\ge 1$ null hypotheses 
\begin{equation*}
H^{(j(n_i,|\mJ_i|))}, H^{(j(n_i,|\mJ_i|-1))}, \ldots, H^{(j(n_i,|\mJ_i|-t_i+1))},
\end{equation*}
where 
\begin{equation*}
t_i=\max\left\{t\in [|\mJ_i|]: \wtilde{\Lambda}^{(j(n_i,\ell))}(n_i)\ge b_{r_i+|\mJ_i|-\ell+1}\qmq{for all}\ell=|\mJ_i|-t+1,\ldots, |\mJ_i| \right\}.
\end{equation*} 

\item If $r_i+t_i=J$ or $n_i=\oN$, terminate the procedure. Otherwise, set $r_{i+1}=r_i+t_i$, let $\mJ_{i+1}$ be the indices of the remaining hypotheses, and continue on to stage $i+1$.
\end{enumerate} 

\subsection{Truncated, rejective procedures controlling FDR or pFDR}
\begin{theorem} \label{thm:fin.fdr} The generic rejective sequential step down procedure implemented with step values $\bm{\alpha}$ satisfies
\begin{equation}\label{fin.fdr.m.bd}
\fdr(\theta)\le D(\bm{\alpha},m_0)\le D(\bm{\alpha}) \qmq{for all}\theta\in\Theta 
\end{equation}
regardless of the dependence between data streams, where $D$ is as defined in \eqref{Dm}-\eqref{D} and $m_0$ is the number of true null hypotheses. In particular, given $q_1\in (0,1)$, the generic rejective sequential step down procedure implemented with step values $\wtilde{\bm{\alpha}}=q_1\bm{\alpha}/D(\bm{\alpha})$ satisfies
\begin{equation}\label{fin.fdr.alpha.bd}
\fdr(\theta)\le q_1 \qmq{for all}\theta\in\Theta 
\end{equation}
regardless of the dependence between data streams.
\end{theorem}

\begin{theorem} \label{thm:fin.pfdr} Let $D$ be as defined in \eqref{Dm}-\eqref{D} and $m_0$ denote the numbers of true null hypotheses. Fix arbitrary $\theta\in\Theta$.
\begin{enumerate}
\item If $\gamma_1>0$ is  such that the generic rejective sequential step down procedure implemented with step values $\bm{\alpha}$ satisfies 
\begin{equation}\label{fin.gamma.lower}
P_\theta(R>0)\ge \gamma_1,
\end{equation}
then this procedure satisfies
\begin{equation}\label{fin.pfdr.m.bd}
\pfdr(\theta) \le \frac{D(\bm{\alpha},m_0)}{P_\theta(R>0)}\le  \frac{D(\bm{\alpha})}{\gamma_1}
\end{equation}
regardless of the dependence between data streams. 

\item\label{part:fin.pfdr.q} In particular, given $q_1\in (0,1)$,  the generic rejective sequential step down procedure implemented with step values $\wtilde{\bm{\alpha}}=q_1\gamma_1\bm{\alpha}/D(\bm{\alpha})$ satisfies
\begin{equation}\label{fin.pfdr.alpha.bd}
\pfdr(\theta)\le q_1
\end{equation}
regardless of the dependence between data streams, if $\gamma_1>0$ satisfies \eqref{gamma.lower} for this procedure.

\item\label{part:fin.pfdr.gij} For $j\in[J]$ let
\begin{equation}\label{fin.g1j.def}
\gamma_{1j}= P_{\theta^{(j)}}(\wtilde{\Lambda}^{(j)}(n)\ge b_1\;\mbox{some $n\le \oN$}).
\end{equation}
If $\max_{j\in[J]}\gamma_{1j}>0$, then \eqref{fin.gamma.lower}-\eqref{fin.pfdr.m.bd} hold with $\gamma_1=\max_{j\in[J]}\gamma_{1j}$.
\end{enumerate}
\end{theorem}

\section{Simulation studies}\label{sec:sims}

In this section we present the results of simulation studies of the
sequential step down procedures described above in Sections~\ref{sec:untrunc} and \ref{sec:trunc} in order to evaluate their operating characteristics
and compare them to analogous fixed-sample procedures. All computations in this section and the analysis of the UK Yellow Card pharmacovigilance database in Section~\ref{sec:Drug-Data} were performed using our Python package available from \url{github.com/bartroff792/mult-seq-dependence}.  First we describe the distributional set ups for our simulation studies, and in then discuss the results in Tables~\ref{tab:sim.binom} and \ref{tab:sim.poiss} and conclude by comparing our proposed sequential procedures with appropriate fixed-sample competitors.

In order to examine the performance of the proposed sequential procedures on correlated Poisson and Binomial data we (i) generated correlated normally distributed data from a Gaussian copula \citep{Trivedi07}, (ii) applied in the inverse Gaussian c.d.f.\ to this data to obtain quantiles, and (iii) applied Poisson and Binomial (respectively) c.d.f.s to the quantiles to obtain correlated data of those marginal distributions. That is, for each time point~$i$  we first
generated a $J$-dimensional multivariate normal vector $Y_i\sim N_J(0,\Sigma)$, using the Toeplitz covariance matrix~$\Sigma_{jj'}=\rho^{|j-j'|}$ for all $j,j'\in[J]$ and where $\rho$ is a chosen value. Then the $J$-dimensional vector of quantiles~$Q_i=\Phi^{-1}(Y_i)$ was computed where $\Phi$ is the standard normal c.d.f.  Finally the data values $X_i^{(j)}$, $j\in[J]$, were set to be
\begin{equation*}
X_i^{(j)}=\begin{cases}
\max\left\{ n:\; F_{\lambda_j}(n)\le Q_{ij}\right\},&\mbox{
for the Poisson case},\\
\bm{1}\{Q_{ij}\le p_j\},&\mbox{for the $\binomial$ case.}
\end{cases}
\end{equation*}
Here $F_\lambda$ is the c.d.f.\ of the Poisson distribution with mean~$\lambda$, and the $\lambda_j$ and $p_j$ are the specified means of the data in the $j$th stream in the Poisson and Binomial cases, respectively.  This process was repeated independently at each needed time point~$i$. The specified value~$\rho$ above equals correlation of ``adjacent'' elements of $Y_i$, corresponding to adjacent data streams, and we note that choosing $\rho<0$, as we do in some of our simulations below, leads to negative dependence among data streams and hence is outside the PRDS condition under which step up and down procedures are already well known \citep[e.g.,][]{Finner09} to control FDR.

For the Binomial data on which the procedures in Table~\ref{tab:sim.binom} were evaluated, individual Bernoulli observations~$X_i^{(j)}$ were generated as described above with mean $p_H=0.05$ under the null and $p_G=0.15$ under the alternative.  For Table~\ref{tab:sim.poiss}, the observations~$X_i^{(j)}$ were generated as Poisson with mean~$\lambda_H=1.5$ under the null, $\lambda_G=2.0$ under the alternative.
For both the Binomial and Poisson data generated to evaluate the procedures in Tables~\ref{tab:sim.binom} and \ref{tab:sim.poiss}, respectively, $J=10$ data streams were utilized scenarios with $m_0=0,1,3,5,7,9,10$ true nulls and $\rho=-.6$ was used in generating copula. Tables~\ref{tab:sim.binom} and \ref{tab:sim.poiss} contain the operating characteristics of the untruncated, FDR and FNR controlling procedures described in Theorem~\ref{thm:inf.fdr}. In both tables, the sequential procedures were implemented with the nominal FDR and FNR error bounds $q_1=.25$ and $q_2=.15$ and using the Wald approximations~\eqref{W4SvS}-\eqref{surr.AB4W} with the null/alternative pairs $\lambda_H,\lambda_G$ and $p_H,p_G$ for the Poisson and Binomial scenarios, respectively, with the BH step values~\eqref{BH.step}.  The  truncated sequential procedures were implemented using Monte Carlo under the known, null parameter values~$\lambda_H, p_H$ to find the critical values, as described in Section~\ref{sec:trunc.setup}.

\begin{table}[ht]
\centering
\caption{Expected sample size~$EN_{Seq}$ and achieved FDR and FNR of sequential step down procedures controlling FDR and FNR, evaluated on negatively correlated Binomial data generated for different numbers~$m_0$ of true null hypotheses; standard errors (SE) are given following each estimated quantity. $N_{FSS}$ is the sample size of the corresponding fixed sample test using the same nominal FDR rate~$q_1=.25$ and which matches the achieved FNR rate of the sequential procedure.}
\label{tab:sim.binom}
\bigskip
\begin{tabular}{c|c|cc|cc|cc}
$m_0$  & $N_{FSS}$& $EN_{Seq}$   & SE   & FDR   & SE    & FNR   & SE    \\\hline
0  & 101 & 36.0 & 0.34 & 0.000 & 0.000 & 0.111 & 0.010 \\
1  & 105 & 39.8 & 0.34 & 0.009 & 0.001 & 0.079 & 0.006 \\
3  & 101 & 45.9 & 0.33 & 0.027 & 0.002 & 0.049 & 0.004 \\
5  & 97  & 50.5 & 0.32 & 0.047 & 0.003 & 0.031 & 0.002 \\
7  & 103 & 53.9 & 0.33 & 0.069 & 0.004 & 0.020 & 0.002 \\
9  & 113 & 55.1 & 0.32 & 0.109 & 0.007 & 0.007 & 0.001 \\
10 & -   & 55.2 & 0.32 & 0.168 & 0.012 & 0.000 & 0.000
\end{tabular}
\end{table}

\begin{table}[ht]
\centering
\caption{Expected sample size~$EN_{Seq}$ and achieved FDR and FNR of sequential step down procedures controlling FDR and FNR, evaluated on negatively correlated Poisson data generated for different numbers~$m_0$ of true null hypotheses; standard errors (SE) are given following each estimated quantity. $N_{FSS}$ is the sample size of the corresponding fixed sample test using the same nominal FDR rate~$q_1=.25$ and which matches the achieved FNR rate of the sequential procedure.}
\label{tab:sim.poiss}
\bigskip
\begin{tabular}{c|c|cc|cc|cc}
$m_0$  & $N_{FSS}$& $EN_{Seq}$   & SE   & FDR   & SE    & FNR   & SE    \\\hline
0  & 83 & 31.6 & 0.22 & 0.000 & 0.000 & 0.107 & 0.010 \\
1  & 79 & 34.3 & 0.23 & 0.009 & 0.001 & 0.077 & 0.006 \\
3  & 77 & 38.1 & 0.24 & 0.029 & 0.002 & 0.052 & 0.004 \\
5  & 73 & 40.4 & 0.25 & 0.050 & 0.003 & 0.038 & 0.003 \\
7  & 79 & 41.6 & 0.26 & 0.077 & 0.004 & 0.023 & 0.002 \\
9  & 99 & 41.0 & 0.26 & 0.119 & 0.008 & 0.007 & 0.001 \\
10 & -  & 40.1 & 0.23 & 0.172 & 0.012 & 0.000 & 0.000
\end{tabular}
\end{table}

The achieved FDR and FNR of the sequential procedures in Tables~\ref{tab:sim.binom} and \ref{tab:sim.poiss} are substantially smaller than their nominal values $q_1=.25$ and $q_2=.15$, although FNR is less so.  To provide a meaningful comparison of these sequential procedures' expected sample size~$EN_{Seq}$ with a comparable fixed-sample size alternative procedure, the sample size~$N_{FSS}$ needed for a fixed-sample BH procedure with the same nominal FDR control level~$q_1=.25$ to match corresponding achieved FNR values are given in the second columns of the tables. 
Thus the sequential procedures provide a large savings in term of sample sizes, at least roughly 45\% savings in all cases and more than 60\% savings in the small $m_0$ cases in both tables, where the adaptivity of the sequential procedures is pronounced. And although the sequential procedures are conservative in the sense of having achieved FDR and FNR values smaller than their nominal values, the corresponding fixed-sample procedures are even more conservative in terms of achieved FDR value due to the larger sample size, while their FNR values were matched for comparison.

\section{Application: The UK's Yellow Card Scheme pharmacovigilance database}\label{sec:Drug-Data}

\subsection{Introduction}
The United Kingdom's Yellow Card pharmacovigilance Database (\url{yellowcard.mhra.gov.uk/information}), run by the Medicines and Healthcare Products Regulatory Agency (MHRA), collects voluntary reports on safety and side-effects from the public and healthcare professionals in the UK on a host of healthcare treatments including medicines, vaccines, medical devices, and even e-cigarettes; in what follows, for simplicity we refer to the treatments in Yellow Card generically as ``drugs.'' The Yellow Card data collection scheme, which now includes a mobile
phone app, began in 1964, spurred by the thalidomide crisis. Its use 
has grown steadily since then, now receiving more than 20,000 reports of possible side effects each year, and totaling more than half a million reports in the scheme’s first 40 years.

\begin{figure}[H]
\centering{}\includegraphics[scale=.6]{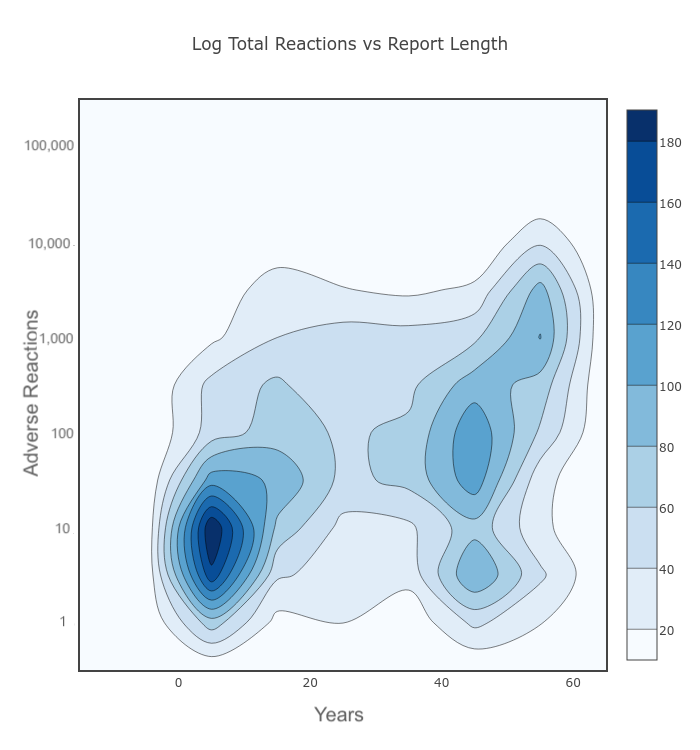}\caption{Number of drugs (indicated by heatmap color) in the Yellow Card database in February~2016 as a function of total adverse reaction reports ($y$ axis, on the $\log_{10}$ scale) and the number of years ($x$ axis) the drug has been in the database.}
\label{fig:drug-database-eda}
\end{figure}

We created a Python script to download and analyze   ``Interactive Drug
Analysis Profiles'' PDF reports from Yellow Card on all of the roughly 2,800 different drugs in the database through February 2016. Figure~\ref{fig:drug-database-eda} is a heatmap showing the number of drugs (indicated by color) in these reports  as a function of their total number of reported adverse action reactions over this period ($y$ axis, on the $\log_{10}$ scale)  and number of years ($x$ axis) the drug has been in the database, calculated from the starting date for collection of reports
on the drug until the closure date of the drug's most recent
summary report.  The figure shows that some of the drugs have many thousands of reaction reports, collected over decades. However, the majority of drug entries have
fewer reports, collected over less than one decade, emphasizing the need for nimble, sequential monitoring of this data.

\subsection{Illustrating the sequential procedures' performance on the database}

Some Yellow Card data has been considered in the statistics literature on multiple testing \citep[e.g.,][]{Dohler18}, but the inherently streaming nature of the data has not been taken into account, with only fixed-sample multiple testing methods being applied to it so far, to our knowledge. In order to demonstrate the behavior of the sequential procedures proposed above on data like that in Yellow Card, we focus on a particular type of side effect report, amnesia, in the Yellow Card reports. 

 The MHRA informed us at the time of our data download that the exact time series of the reports to Yellow Card were not being made available, so it is not possible to apply our sequential procedures to the data precisely as it was being received by Yellow Card.  But, for each of the drug reports downloaded as described above and illustrated in Figure~\ref{fig:drug-database-eda}, we were able to obtain the average number of amnesia and  other side effect reports
per year,  the starting date for collection of reports
for each drug, and the date of the closure of the drug's most recent
summary report.  So here we apply our sequential procedures to data streams simulated using the actual yearly average rates obtained from Yellow Card.  This  approach, similar to the parametric Bootstrap  \citep{Efron94}, makes our simulation as close as possible to real Yellow Card data, given the limitations in data availability.

\begin{figure}[H]
\includegraphics[scale=0.5]{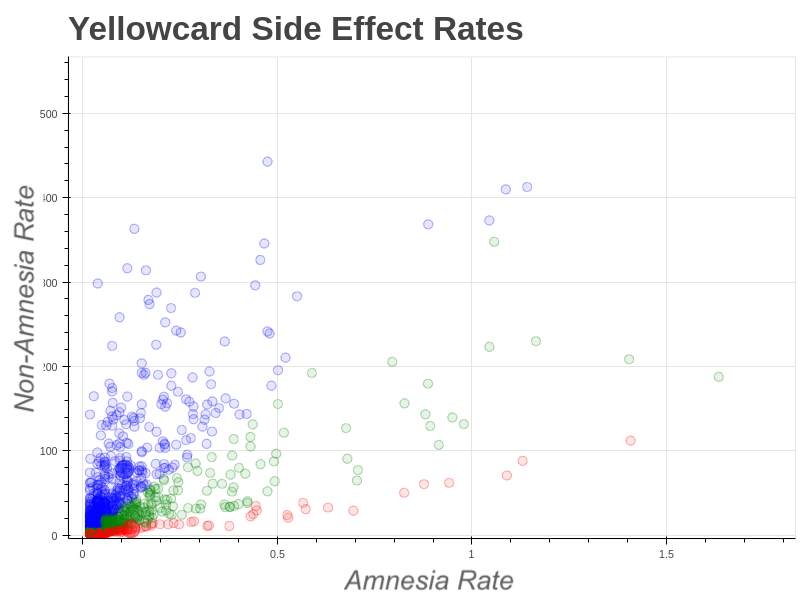}\caption{Drug side effect rates~\eqref{eq:drug-amnesia-rate}-\eqref{eq:drug-nonamnesia-rate} from the Yellow Card database data. Blue points are those drugs~$j$ with $p^{(j)}\leq p_H$, green those with $p_H<p^{(j)}<p_G$, and
red those with $p^{(j)}\geq p_G$, as defined by \eqref{YC.p.binom} and \eqref{eq:drug-p0}. }
\label{fig:Drug-side-effect-raw}
\end{figure}

The Yellow Card data was used to simulate future drug reports as follows. At each time step a Poisson number of amnesia and
non-amnesia reports for each drug was generated as correlated Poisson processes with rates 
\begin{align}
\lambda_{amn}^{(j)} & =\left(|\left\{ \text{amnesia reports from drug~$j$}\right\} |+1\right)/T_j\label{eq:drug-amnesia-rate}\\
\lambda_{-amn}^{(j)} & =\left(|\left\{ \text{non-amnesia reports from drug~$j$}\right\} |+1\right)/T_j,\label{eq:drug-nonamnesia-rate}\qm{where}\\
T_j&=(\mbox{drug~$j$ most recent report}) - (\mbox{drug~$j$ start date}).\nonumber
\end{align}
Note that we have added $1$ amnesia and non-amnesia report to each drug in \eqref{eq:drug-amnesia-rate}-\eqref{eq:drug-nonamnesia-rate} to account for issues with rare or exotic drugs like inherently higher variance in use and reporting.  These rates are visualized in Figure~\ref{fig:Drug-side-effect-raw}, where the blue, green, and red points are the drugs with ``low,'' ``medium,'' and ``high'' amnesia report rates according to \eqref{YC.p.binom} and\eqref{eq:drug-p0}. 

In order to generate a medically realistic correlation structure for the drugs, we utilized BioGPT \citep{Luo22}, a generative language model trained on biomedical research articles.  For each drug under consideration, we calculated the element-wise mean of the BioGPT's \textit{token embedding} (its numerical representation of phrases, words, or characters)
for ``[DrugName] as a drug'' and then applied $K$-means clustering to obtain $K=15$ clusters of the drugs.  Within each cluster, the copula method described in Section~\ref{sec:sims} was then used to generate the Poisson data using the Toeplitz covariance matrix given there, but with the correlation parameter~$\rho$ generated as a Beta$(4,2)$ random variable, shifted and scaled to have support~$[-1,1]$.  In this way, drugs that were medically related according to BioGPT were more likely to receive non-trivial correlation.

In order to account accurately for the differing number of reports that drugs with different prescription and use rates in the database have, we set up the hypothesis tests not in terms of the Poisson rate parameters~\eqref{eq:drug-amnesia-rate}-\eqref{eq:drug-nonamnesia-rate}
but rather a scaled parameter
\begin{equation}\label{YC.p.binom}
p^{(j)}=\frac{\lambda_{amn}^{(j)}}{\lambda_{amn}^{(j)}+\lambda_{-amn}^{(j)}}\approx \frac{\mbox{\# amnesia reports  for drug~$j$}}{\mbox{total \# reports for drug~$j$}}.
\end{equation} The data generation scheme above is equivalent to a Binomial number of amnesia reports being generated at each time step, with success probability~\eqref{YC.p.binom} and number of trials equal to the Poisson random total number of  reports.

To set up our hypothesis tests, we utilize the Yellow Card database to calculate 
\begin{equation}
p_H  = \mbox{50th percentile of}\; \{p^{(j)}\},\qmq{and} p_G  = \mbox{90th percentile of}\; \{p^{(j)}\}\label{eq:drug-p0}
\end{equation}
which we use as surrogates for ``typical'' and ``extreme'' Binomial rates of amnesia reports, respectively. After calculating $p_H$ and $p_G$ on the entire Yellow Card database, we then applied out sequential procedure to the $J=1800$ drugs with the highest total side effect reports in the database.  This reduction still maintains a large number of data streams but also allows us to filter out drugs that have only recently been made available with too small a number of reports, but also the opposite extreme of filtering out obscure and antiquated drugs that exist in the database (e.g., ``Mammalian Blood'' and ``Wool
Fat''). 

For each of these $J=1800$ drugs we tested the hypotheses
$$H^{(j)}:p^{(j)}\le p_H\qmq{vs.} G^{(j)}:p^{(j)}\ge p_G>p_H$$
representing that the drug has a ``typical'' versus abnormally high rate of amnesia reports. For these we implemented the open-ended sequential test described in Theorem~\ref{thm:inf.fdr} for Binomial data using FDR controlled at $q_1=0.05$ and FNR at $q_2=0.15$, using with the Wald approximations~\eqref{W4SvS}-\eqref{surr.AB4W} the BH step values~\eqref{BH.step}. Running this data generation scheme with $1800$ data streams a single time is computationally intensive even on a Google Cloud Computing instance, so rather than report Monte Carlo statistics, in Table~\ref{tab:YC} we give the output of a typical run and there list the first 15 drugs whose streams terminated to accept, and the first 15 to reject, the null hypothesis~$H^{(j)}$.  Since the simulation is set up where each ``observation'' is the number of reports in a year, in the table the Termination Step corresponds to the number of years required for termination of each stream.  The Termination Level is a measure of the strength of the terminal action, where smaller values indicate a stronger acceptance/rejection of the null, and equals the \emph{cumulative} value of the level~$\ell$ in step~\ref{rej-step} or \ref{acc-step} of the procedure as defined in Section~\ref{sec:gen.D}.

\subsection{Discussion}
While we draw no medical conclusions from our simulations, we note that some of the drugs identified by our sequential procedure by stopping  to reject their null hypotheses in favor of an abnormally high level of amnesia reports, have been associated with amnesia in the biomedical literature \citep[e.g.,][]{Chavant11}.  Monitoring pharmacovigilance databases for amnesia and other cognitive impairment adverse events is  a real-world priority of statisticians and regulators around the world; see \citet{Borchert19} and \citet{Sato20} for examples of such analyses in the U.S.'s Food and Drug Administration Adverse Event Reporting System (FAERS; see \url{www.fda.gov/media/97567/download}), and \citet{Sunwoo24} for an analysis of the Korean Adverse Event Reporting System Database.  The example presented here provides a possible new statistical tool for this important task that takes into account the real-time, streaming  nature of the data as well as the inherent multiple comparisons issues with such data. Although we are aware of proposed approaches that separately take into account the time-sequential nature of the data \citep[e.g.,][]{Hocine09} and the multiple comparison aspect, this is the first proposal to incorporate both that we are aware of.

The dominant statistical methods in this existing literature are ``disproportionality analyses'' \citep[see][]{Bihan20,Lavertu21,Zhang24,Zorych13}, utilizing odds (or log odds) ratios of the probability of reporting an adverse event, which is equivalent to our use of the probabilities of such reports in \eqref{YC.p.binom}-\eqref{eq:drug-p0}.  There we use the  probabilities of such reports across the entire database as the levels of ``typical'' and ``extreme'' in \eqref{eq:drug-p0} to set our hypotheses. This approach, known as ``signal of disproportional reporting,'' is common in the literature \citep[e.g.,][]{Hauben05,Hauben05b,Zorych13}, although different approaches exist on how to choose these cutoffs and some of these authors use individual-level covariates, such as age in \citet{Hocine09}, in setting these cutoffs, which was not available in our example. However, signal of disproportional reporting is not a requirement of our method proposed here and those levels could be set by other methods such as from the medical literature, regulatory concerns, etc.

The approach proposed here could be used with other types of drugs and side effects, although domain-specific details the particular application will likely affect the choices made in  implementation.  For example, in monitoring adverse side effects of statins in the Korean pharmacovigilance database, \citet{Kim17} grouped together statin-specific symptoms and measured severity on the WHO-specified scale, and the life-threatening nature of adverse immune-mediated reactions associated with certain immunotherapies \citep[e.g.,][]{Ali17} would be taken into account in setting the early stopping properties of the procedure.

\begin{table}[ht]
\centering\caption{The earliest 15 Yellow Card database drugs to terminate to reject, or accept, the null hypothesis~$H^{(j)}$ in a single run of the sequential step down procedure controlling FDR at $q_1=.05$ and FNR at $q_2=.15$. The Termination Step corresponds to the number of years, and termination level is the cumulative value of the level~$\ell$ in step~\ref{rej-step} or \ref{acc-step} of the procedure as defined in Section~\ref{sec:gen.D}. cAC10-vcMMAE is an abbreviation for \textit{Chimeric Human Murine Monoclonal Antibody Cac10 Anti Cd30 Linked To Cytotoxic Molecule Sgd 1006}.}
\bigskip
\begin{tabular}{lccc}
Drug               & Terminal Action & Termination Step & Termination Level \\\hline\hline
Bupropion          & Accept $H^{(j)}$        & 1                & 2                 \\
Clozapine          & Accept $H^{(j)}$        & 1                & 2                 \\
Etanercept         & Accept $H^{(j)}$        & 2                & 8                 \\
Venlafaxine        & Accept $H^{(j)}$        & 2                & 8                 \\
Varenicline        & Accept $H^{(j)}$        & 2                & 8                 \\
Paroxetine         & Accept $H^{(j)}$        & 2                & 8                 \\
Clostridium Tetani & Accept $H^{(j)}$        & 2                & 8                 \\
Telaprevir         & Accept $H^{(j)}$        & 2                & 8                 \\
Adalimumab         & Accept $H^{(j)}$        & 3                & 10                \\
Rofecoxib          & Accept $H^{(j)}$        & 3                & 10                \\
Ranibizumab        & Accept $H^{(j)}$        & 4                & 21                \\
Amoxycillin        & Accept $H^{(j)}$        & 4                & 21                \\
Oseltamivir        & Accept $H^{(j)}$        & 4                & 21                \\
Trimethoprim       & Accept $H^{(j)}$        & 4                & 21                \\
Ciprofloxacin      & Accept $H^{(j)}$        & 4                & 21                \\\hline
Zopiclone          & Reject $H^{(j)}$        & 7                & 1                 \\
Simvastatin        & Reject $H^{(j)}$        & 9                & 2                 \\
Zolpidem           & Reject $H^{(j)}$        & 13               & 3                 \\
Bifonazole         & Reject $H^{(j)}$        & 15               & 5                 \\
Rimonabant         & Reject $H^{(j)}$        & 15               & 5                 \\
Enzalutamide       & Reject $H^{(j)}$        & 25               & 7                 \\
Gabapentin         & Reject $H^{(j)}$        & 25               & 7                 \\
Atorvastatin       & Reject $H^{(j)}$        & 31               & 8                 \\
cAC10-vcMMAE       & Reject $H^{(j)}$        & 32               & 9                 \\
Dutasteride        & Reject $H^{(j)}$        & 33               & 10                \\
Mitotane           & Reject $H^{(j)}$        & 36               & 11                \\
Boceprevir         & Reject $H^{(j)}$        & 39               & 12                \\
Lurasidone         & Reject $H^{(j)}$        & 42               & 14                \\
Asenapine          & Reject $H^{(j)}$        & 42               & 14                \\
Retigabine         & Reject $H^{(j)}$        & 44               & 16               
\end{tabular}
\label{tab:YC}
\end{table}

\section{Conclusion}\label{sec:conc}
We have proposed both open-ended and truncated sequential multiple testing procedures which can control FDR or pFDR (and their type~II analogs, FNR or pFNR in the open-ended case) under arbitrary dependence between the data streams. These procedures have shown large savings in average sample size compared to their fixed-sample counterparts in our simulation studies, and in Section~\ref{sec:Drug-Data} we demonstrate how these procedures may be used to monitor data streams like those coming from a pharmacovigilance database, like the UK's Yellow Card database.

\section*{Compliance with Ethical Standards} The authors have no conflicts of interest to report.

%\bibliographystyle{apalike}
%\bibliography{../../Bib_files/bibliography}

\def\cprime{$'$}

\section*{Appendix}
\appendix
\section{Proof of Theorem~\ref{thm:inf.fdr}}\label{sec:inf.pf}
We first prove the inequalities for FDR in \eqref{inf.fdr.m.bd} for the sequential step down procedure with step values $\bm{\alpha}, \bm{\beta}$, and then discuss those for FNR.  The inequalities in \eqref{inf.fdr.alpha.bd} then follow, using \eqref{D.lin}. Our proof largely follows the arguments of \citet{Guo08}, with a few important differences to account for the sequential nature of our procedures. 

Fix arbitrary $\theta$ and omit it from the notation, and below let $P(\cdot)$ and $E(\cdot)$ denote the probability and expectation under $\theta$ and an arbitrary joint distribution between the data streams. Without loss of generality assume $H^{(1)},\ldots, H^{(m_0)}$ are the true hypotheses for some $m_0\in[J]$, the $m_0=0$ case being trivial since $\fdr=0$ in this case. For $i\in[J]$ let $\tau_i$ denote the time at which the $i$th stream~$\wtilde{\Lambda}^{(i)}$ terminates, and for $i,j\in[J]$ let
\begin{equation}\label{inf.Fij.def}
F_{ij}=\{\wtilde{\Lambda}^{(i)}(\tau_i)\in [b_j, b_{j-1})\},
\end{equation}
setting $b_0=\infty$ to handle the $j=1$ case.  The event $F_{ij}$ is similar (in fact, contained in) the event in \eqref{typeI.stand}, but it specifies which interval~$[b_j, b_{j-1})$ the $i$th statistic~$\wtilde{\Lambda}^{(i)}$ is in when stopping to reject its corresponding null~$H^{(i)}$; we will refer to this as \textit{$H^{(i)}$ being rejected at the $j$th level}. Recalling that  $R$ and $V$ denote the number of null and true null hypotheses, respectively, rejected by the procedure, define 
\begin{equation}\label{pijk.def.inf.fdr}
p_{ijk} = P(F_{ij}\cap \{R=k\})\qmq{for} i,j, k\in[J],
\end{equation} 
with which we can write
\begin{equation}\label{fdr.pijk}
\fdr = \sum_{i=1}^{m_0}\sum_{j=1}^J \sum_{k=j}^J \frac{p_{ijk}}{k},
\end{equation} which is analogous to similar expressions obtained for FDR in the fixed-sample setting obtained by \citet{Benjamini01} and \citet{Sarkar02}. To see why \eqref{fdr.pijk} holds here for the sequential procedure, write
\begin{multline*}
\fdr = E\left(\frac{V}{R\vee 1}\right) = \sum_{k=1}^J \frac{1}{k}E(V\bm{1}\{R=k\}) = \sum_{k=1}^J\sum_{i=1}^{m_0} \frac{1}{k} P(\{\mbox{$H^{(i)}$ rejected}\}\cap \{R=k\}) \\
= \sum_{k=1}^J\sum_{i=1}^{m_0} \sum_{j=1}^k\frac{1}{k} P(F_{ij}\cap\{R=k\})= \sum_{i=1}^{m_0}\sum_{j=1}^J\sum_{k=j}^J \frac{1}{k} p_{ijk},
\end{multline*}
 where in the last equality we have reordered the sums and used the definition of $p_{ijk}$.
 
 We further extend the notation to expand \eqref{fdr.pijk}. Given $k\in[J]$, $\ell\in[k\wedge m_0]$, and a pair of vectors~$(\bm{i}, \bm{j})$  in 
 \begin{equation*}
\Omega_{\ell k}=\{(\bm{i},\bm{j})\in [m_0]^\ell\times [k]^\ell: 1\le i_i<i_2<\ldots<i_\ell\le m_0\},
\end{equation*} define
 \begin{equation}\label{p.bold.def}
p_{\bm{i}\bm{j}k} = P\left(\bigcap_{d=1}^\ell F_{i_d j_d} \cap \{R=k\} \cap \{V=\ell\}\right),
\end{equation} which is the probability of $k$ rejections, of which the false rejections are $H^{(i_d)}$ being rejected at the $j_d$th level, $d\in[\ell]$. Let  
\begin{equation*}
\Omega_{\ell k}(i,j)=\{(\bm{i},\bm{j})\in \Omega_{\ell k}: (i,j)=(i_d,j_d)\;\mbox{for some $d\in[\ell]$}\}
\end{equation*} be the set of those vector pairs that include $H^{(i)}$ being rejected at the $j$th level.  With these definitions we can further expand $p_{ijk}$ in the form
\begin{equation}\label{p=pbm}
p_{ijk} = \sum_{\ell=1}^{k\wedge m_0} \sum_{(\bm{i},\bm{j})\in\Omega_{\ell k}(i,j)} p_{\bm{i}\bm{j}k}.
\end{equation} This equality holds because, for distinct pairs $(\bm{i},\bm{j})\in\Omega_{\ell k}(i,j)$, the events in \eqref{p.bold.def} correspond to different nulls being rejected, or at different levels, and are hence disjoint, thus
\begin{equation*}
\sum_{(\bm{i},\bm{j})\in\Omega_{\ell k}(i,j)} p_{\bm{i}\bm{j}k} = P\left( F_{ij} \cap \{R=k\} \cap \{V=\ell\}\right).
\end{equation*} For distinct values of $\ell$, the events in this last are clearly disjoint, hence summing over $\ell$ yields
\begin{equation*}
\sum_{\ell=1}^{k\wedge m_0} \sum_{(\bm{i},\bm{j})\in\Omega_{\ell k}(i,j)} p_{\bm{i}\bm{j}k} = \sum_{\ell=1}^{k\wedge m_0}P\left( F_{ij} \cap \{R=k\} \cap \{V=\ell\}\right) = P(F_{ij}\cap \{R=k\}) = p_{ijk}.
\end{equation*}
With \eqref{p=pbm} established we can substitute it into \eqref{fdr.pijk} to obtain
\begin{multline}\label{fdr.pbold}
\fdr = \sum_{i=1}^{m_0}\sum_{j=1}^J \sum_{k=j}^J \sum_{\ell=1}^{k\wedge m_0} \sum_{(\bm{i},\bm{j})\in\Omega_{\ell k}(i,j)}  \frac{p_{\bm{i}\bm{j}k}}{k} =  \sum_{\ell=1}^{m_0}\sum_{k=\ell}^J\sum_{i=1}^{m_0}\sum_{j=1}^k \sum_{(\bm{i},\bm{j})\in\Omega_{\ell k}(i,j)}  \frac{p_{\bm{i}\bm{j}k}}{k}\\
=  \sum_{\ell=1}^{m_0}\sum_{k=\ell}^J \sum_{(\bm{i},\bm{j})\in\Omega_{\ell k}}  \frac{\ell p_{\bm{i}\bm{j}k}}{k}.
\end{multline} The second equality in \eqref{fdr.pbold} is obtained by reordering the summations, and the last equality uses
\begin{equation*}
\sum_{i=1}^{m_0}\sum_{j=1}^k \sum_{(\bm{i},\bm{j})\in\Omega_{\ell k}(i,j)}  \frac{p_{\bm{i}\bm{j}k}}{k} = \sum_{(\bm{i},\bm{j})\in\Omega_{\ell k}} \frac{\ell p_{\bm{i}\bm{j}k}}{k},
\end{equation*} 
since each pair $(\bm{i},\bm{j})\in\Omega_{\ell k}$ appears in the $\ell$ sets $\Omega_{\ell k}(i_1,j_1),\ldots, \Omega_{\ell k}(i_\ell,j_\ell)$, and only those sets.

With the expression \eqref{fdr.pbold} for FDR established, we now consider constraints that the $p_{\bm{i}\bm{j}k}$ must satisfy.  For the first constraint, take arbitrary $k\in[J]$, $\ell\in[k\wedge m_0]$, $(\bm{i},\bm{j})\in\Omega_{\ell k}$, and let $j_{(1)}\le\ldots\le j_{(\ell)}$ denote  an ordering of the values in $\bm{j}=(j_1,\ldots,j_\ell)$.  For an arbitrary  fixed-sample step down procedure with $p_{\bm{i}\bm{j}k}$ defined analogously, \citet[][Lemma~3.3]{Guo08} show that
\begin{equation}\label{pbm=0}
p_{\bm{i}\bm{j}k}=0\qmq{if} k-\ell> m-m_0\qmq{or} j_{(d)}>k-\ell+d\;\mbox{for some $d\in[\ell]$.}
\end{equation} Their proof depends only on the step down structure and the values that  $R, V$ can take with positive probability, and thus hold for our sequential step down procedure as well, so we do not repeat the proof here. 

The other constraint relates the $p_{\bm{i}\bm{j}k}$ to the error probabilities in \eqref{typeI.stand}. First observe that, for any $i\in[m_0]$, the events $F_{i1}, F_{i2},\ldots$ are disjoint, as are $\{R=1\}, \{R=2\}, \ldots$.  Then,  for any $s\in[J]$, we write \eqref{typeI.stand} as
\begin{multline}\label{alpha>pbm}
\alpha_s\ge P\left(\bigcup_{j=1}^s F_{ij}\right) = \sum_{j=1}^s P(F_{ij}) = \sum_{j=1}^s P\left(F_{ij} \cap \bigcup_{k=1}^J  \{R=k\} \right) = \sum_{j=1}^s \sum_{k=1}^J P\left(F_{ij} \cap \{R=k\} \right)\\
= \sum_{j=1}^s \sum_{k=1}^J p_{ijk} = \sum_{j=1}^s \sum_{k=1}^J\sum_{\ell=1}^{k\wedge m_0} \sum_{(\bm{i},\bm{j})\in\Omega_{\ell k}(i,j)} p_{\bm{i}\bm{j}k},
\end{multline} using \eqref{p=pbm} for this last equality.

Combining \eqref{fdr.pbold}, \eqref{pbm=0}, and \eqref{alpha>pbm}, the goal of finding the worst-case joint distribution that maximizes FDR can be stated as finding the $\{p_{\bm{i}\bm{j}k}\}$ that
\begin{equation}\label{opt.prob}
\mbox{maximize}\; \fdr =\sum_{\ell=1}^{m_0}\sum_{k=\ell}^J \sum_{(\bm{i},\bm{j})\in\Omega_{\ell k}}  \frac{\ell p_{\bm{i}\bm{j}k}}{k}\qm{subject to \eqref{pbm=0} and \eqref{alpha>pbm}.} 
\end{equation}
At this point the $\{p_{\bm{i}\bm{j}k}\}$ can be completely divorced from their original meaning about a multiple testing procedure and treated as arbitrary variables in the constrained optimization problem~\eqref{opt.prob}.  \citet{Guo08} solve a similar problem, identical to \eqref{opt.prob} except for the second constraint; their second constraint is sufficient but not necessary for our second constraint~\eqref{alpha>pbm}; about this, see Remark~\ref{rem:GR.valid}. Guo and Rao proceed to solve this problem by producing an upper bound on FDR that coincides with FDR at its maximum~$D(\bm{\alpha}, m_0)$, and then show that the maximizer satisfies both constraints.  Since their second constraint is sufficient for ours, their proof applies here, and we do not repeat more of the details here. This establishes the inequalities for FDR in \eqref{inf.fdr.m.bd}.

The inequalities for FNR in \eqref{inf.fdr.m.bd} are established in a completely similar way after reversing the roles of the ``type~I'' and ``type~II'' objects, e.g., substituting FNR for FDR, $\bm{\beta}$ for $\bm{\alpha}$, the bound \eqref{typeII.stand} for \eqref{typeI.stand}, etc. The details are straightforward and therefore omitted here. \qed

\begin{remark}\label{rem:GR.valid} Guo and Rao's \citeyearpar[][expression (19)]{Guo08} second constraint involves bounding from above the probability that a $p$-value for a true null falls in the interval $[\alpha_{j-1},\alpha_j)$ by $\alpha_j-\alpha_{j-1}$. Validity of the $p$-value is not sufficient for this to hold and, for example, it may fail  for valid but discrete $p$-values. Therefore their proof actually requires a condition stronger than validity on the $p$-values, for example having the exact uniform $(0,1)$ null distribution. We avoid the need for such a stronger condition by summing over $j$ in \eqref{alpha>pbm}, and thus only need \eqref{typeI.stand} which is the sequential analog of validity for $p$-values in the fixed sample set up.
\end{remark}

\section{Sketch of proof of Theorem~\ref{thm:inf.pfdr}}
The proof of the inequalities for $\pfdr{}$ in \eqref{inf.pfdr.m.bd} is similar to the proof of Theorem~\ref{thm:inf.fdr} after replacing $p_{ijk}$ in \eqref{pijk.def.inf.fdr} by 
\begin{equation*}
p_{ijk} = P(F_{ij}\cap \{R=k\}|R>0)
\end{equation*} 
and using \eqref{gamma.lower} to bound $P(R>0)$ from below, leading to the factor of $1/\gamma_1$ in the bound.  The proof of the inequalities for $\pfnr{}$ in \eqref{inf.pfdr.m.bd} proceeds similarly, conditioning on $R<J$ and using $\gamma_2$. Part~\ref{part:inf.pfdr.q} of Theorem~\ref{thm:inf.pfdr} then follows using \eqref{D.lin}.  In Part~\ref{part:inf.pfdr.gij}, the event in \eqref{g1j.def} implies rejection of $H^{(j)}$, hence $\max_j \gamma_{1j}\le P(R>0)$, with analogous statements applying to the type~II version.

\section{Sketch of proofs of Theorems~\ref{thm:fin.fdr} and \ref{thm:fin.pfdr} }
The proof of Theorem~\ref{thm:fin.fdr} follows the proof of FDR control in Theorem~\ref{thm:inf.fdr} with $p_{ijk}$ defined in \eqref{pijk.def.inf.fdr} but with
\begin{equation}\label{fin.Fij.def}
F_{ij}=\{\wtilde{\Lambda}^{(i)}(\tau_i)\in [b_j, b_{j-1}),\quad \tau_i<\oN\}
\end{equation}
rather than \eqref{inf.Fij.def}.

For Theorem~\ref{thm:fin.pfdr}, the proof proceeds the similarly but with $p_{ijk}$ replaced by
\begin{equation*}
p_{ijk} = P(F_{ij}\cap \{R=k\}|R>0),
\end{equation*} 
again with $F_{ij}$ given by \eqref{fin.Fij.def}. Then \eqref{fin.gamma.lower} is used to bound $P(R>0)$ from below, giving the factor of $1/\gamma_1$ in the bound.  Part~\ref{part:fin.pfdr.q} follows from \eqref{D.lin}, and the event in \eqref{fin.g1j.def} of Part~\ref{part:inf.pfdr.gij} implies rejection of $H^{(j)}$, hence $\max_j \gamma_{1j}\le P(R>0)$.

\end{document}